\documentclass[%
 reprint,
 amsmath,amssymb,
 aps,
 pra,
 floatfix,
]{revtex4-1}
\usepackage{verbatim}
\usepackage{algorithm}
\usepackage{natbib}
\usepackage[most]{tcolorbox}
\usepackage[noend]{algpseudocode}
\usepackage{graphicx}
\usepackage{dcolumn}
\usepackage{bm}
\DeclareMathOperator*{\argmin}{arg\,min}
\usepackage{float}

\begin{document}


\title{Box algorithm for the solution of differential equations on a quantum annealer}

\author{Siddhartha Srivastava}
\email{PhD candidate, sidsriva@umich.edu}
\author{Veera Sundararaghavan}%
 \email{Associate Professor, veeras@umich.edu}
\affiliation{%
Department of Aerospace Engineering\\
University of Michigan Ann Arbor MI 48109\\
}%

\date{\today}

\begin{abstract}
Differential equations are ubiquitous in models of physical phenomena. Applications like steady-state analysis of heat flow and deflection in elastic bars often admit to a second order differential equation. In this paper, we discuss the use of a quantum annealer to solve such differential equations by recasting a finite element model in the form of an Ising hamiltonian. The discrete variables involved in the Ising model introduce complications when defining differential quantities, for instance, gradients involved in scientific computations of solid and fluid mechanics. To address this issue, a graph coloring based methodology is proposed which searches iteratively for solutions in a subspace of weak solutions defined over a graph, hereafter called as the `box algorithm.' The box algorithm is demonstrated by solving a truss mechanics problem on the D-Wave quantum computer.
\end{abstract}

\pacs{Valid PACS appear here}
\maketitle

\section{Introduction}

Computational methods are rapidly emerging as an essential tool to understand and solve complex engineering problems complementing the traditional means of experimentation and theory.  Richard Feynman's statement \cite{feynman1982simulating} ``with a suitable class of quantum machines you could imitate any quantum system, including the physical world" has driven our vision towards a machine that can solve computational problems inaccessible to classical computers. Early versions of such quantum computers have already appeared. Mirroring gate-based classical computers, gate--based quantum computers with a small number of qubits have been demonstrated and promise an eventual path towards universal quantum computation. However, noise limits the number of gate operations that can be enforced before the quantum states decohere. In parallel, quantum annealers have been developed that provide a significant number of qubits for solving a class of combinatorial optimization problems. In these machines, an Ising hamiltonian is engineered such that the solution to the computational problem is encoded in its ground state. The system evolves adiabatically to the ground state as governed by the Schrodinger equation for the time-dependent hamiltonian. 

The D-wave system is a quantum annealer that currently provides more than 2000 qubits modeling a transverse Ising hamiltonian whose ground state is NP-complete. The  hamiltonian with $q_i$ describing the state of the $i^{th}$ qubit is given by: 
\begin{equation}\label{eq_Ising}
    E(\mathbf{q}) = \sum_{i \in \textrm{sites}} H_i q_i + \sum_{(i,j) \in \textrm{links}} J_{ij} q_i q_j 
\end{equation}
The hamiltonian includes self-interaction ($H_i$ is the on-site energy of qubit $q_i$) and site-site interaction terms ($J_{ij}$ are the interaction energies of two connected qubits $q_i$ and $q_j$), with the qubits connected in a Chimera graph.  The system is first initialized in the ground state of a hamiltonian which is known and easy to prepare. 
Then, the Hamiltonian is changed such that the system equilibrates to the ground state of the Ising Hamiltonian $E(\mathbf{q})$ according to the adiabatic theorem.
NP-hard combinatorial optimization problems can be encoded through the field and site interaction strengths. The system particularly holds promise for solving graph coloring problems with large sizes (N) where classical polynomial time algorithms cannot be devised.   Many engineering problems in airline scheduling, image segmentation, and pattern recognition have been encoded as graph coloring problems solvable on quantum annealers.  

While differential equations are ubiquitous in models of physical phenomena, the use of quantum annealers for scientific computing in solid and fluid mechanics has not yet been explored. Scientific computing mostly involves solving a linear system of equations $Ax=b$ defined on a continuum domain discretized with finite elements. The matrix $A$, generally being sparse, structured and positive definite matrix obtained by assembling element-level stiffness matrices. In the past, gate--based quantum computing algorithms have been devised to solve the system of linear equations using QLSA algorithms (HHL algorithm \cite{harrow2009quantum}) and its variants \cite{ambainis2010variable,childs2015quantum,clader2013preconditioned,wossnig2018quantum}. This algorithm, unlike a classical solver, does not give a direct solution $x$ but rather allows sampling from the solution vector. Nevertheless, this has spawned several works in differential equation modeling on quantum computers (\cite{berry2010quantum,leyton2008quantum,montanaro2016quantum,cao2013quantum,pan2014experimental,steijl2018parallel,costa2017quantum,fillion2017simple,sun2017solving}). The sampling task by itself requires solving  $Ax=b$. In the classical setting, the complexity scales with the size of the problem and goes as $O(Nsk\log(1/\epsilon))$ for conjugate gradient method where $N$ is the number of unknowns, $k$ is the condition number, $s$ is the sparsity of $A$ and $\epsilon$ is the precision of the solution. On the other hand, the QLSA \cite{harrow2009quantum} has a favorable running time of $O(\log(N)k^2s^2/\epsilon)$ which scales logarithmically with the size of the problem. Quantum annealers are especially attractive for scientific computing with the ability to scale up the simulations to a more significant number of qubits. However, algorithms for the solution of differential equations have not been devised yet on these systems\cite{lanlreport}. The merit of the paper is the algorithm to solve a differential equation on an annealer. 

\begin{figure*}[hbt!]
\centering
\includegraphics[width=0.8\textwidth]{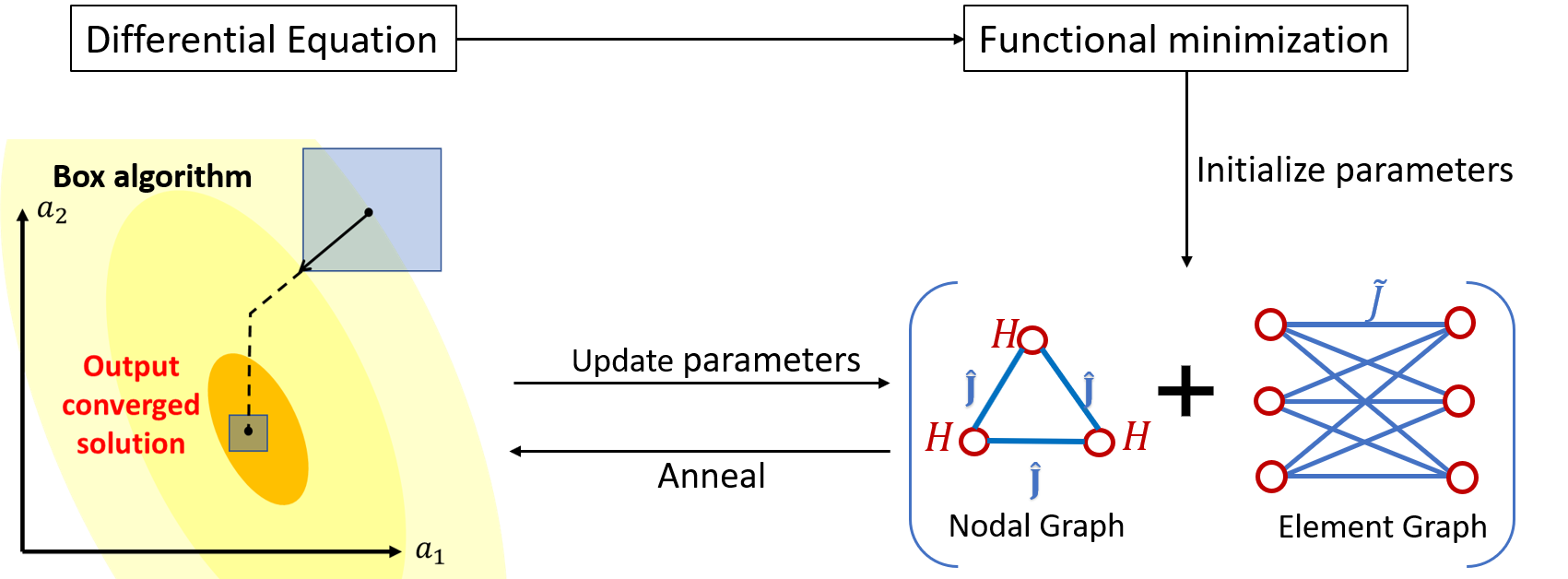}
\caption{Illustration of procedure for solving differential equation}
\label{fig_process}
\end{figure*}

Here, we note that the solution to $Ax=b$ can be encoded in an equivalent minimization problem $min \left(\frac{1}{2} x^TAx - x^Tb\right)$ which contains field and interaction terms similar to an Ising model. Thus, in this paper, we explore mapping of this energy to Ising hamiltonian on the D-wave machine recognizing that the graph in the D-wave chip by itself models a finite element mesh-like topology. The element level stiffness and force vectors are then encoded in the Ising hamiltonian as interaction weights and field variables. Dirichlet boundary conditions are enforced by modifying field terms to favor one qubit state over another. An illustration of this procedure is presented in Fig \ref{fig_process}. A discretized version of the differential equation is solved using energy minimization on a graph. Direct minimization of energy may hold advantages over the conventional finite element approach in systems which lead to a matrix with large condition numbers, zero or negative eigenvalues leading to bifurcation events such as buckling in shells and phase transitions \cite{ba2010stability}.\\

In the solution to a differential equation, the qubits must encode a rational number. However, the qubit encoding the Ising lattice point carries two discrete levels (up/down spin) in the ground state. In classical computers, with similar binary (0/1) encoding, anywhere from 32 bits (float) to 80 bits (long double) of memory can be used to encode more than 12 million high precision variables in 1 GB memory. In contrast, currently available quantum annealers have a limited number of physical qubits. This restriction makes the representation of solutions of double precision similar to a classical computer extremely expensive.
In (\cite{borle2018analyzing,o2018nonnegative}), the problem of minimizing $||Ax-b||$ in the least squares sense was posed by encoding physical qubits to represent rational numbers using a radix 2 representation. This format requires a significant number of physical qubits and connections to represent positive rational numbers and an additional qubit to represent the sign of the number (\cite{borle2018analyzing}). 
In comparison, the box algorithm searches within a small discrete set of up/down qubit values with each element of the set mapped to a double precision value, thereby eliminating the need for additional qubits to achieve higher precision. 

In this paper, we consider a self-adjoint form of second order differential equation as the model problem. The problem statement and the relevant mathematical details are presented in Section \ref{sec_math}. The Graph representation of the problem is formulated in Section \ref{sec_Graph_model}.  The iterative procedure, referred as 'Box algorithm', is presented in Section \ref{sec_Box}. All procedures are accompanied with numerical examples for elucidation. This algorithm is demonstrated by solving a truss mechanics problem on the D-Wave quantum computer in Section \ref{sec_results}

\section{Mathematical Preliminaries}\label{sec_math}

A self-adjoint form of a second order differential equation on an interval $(x_l, x_r)$ is defined as, 
\begin{eqnarray}\label{eq_strongform}
-(p(x)u'(x))' + q(x)u(x) = f(x) \qquad x_l<x<x_r
\end{eqnarray}
Dirichlet boundary conditions are considered at both ends i.e.
$u(x_l) = u_l \textrm{ and }u(x_r) = u_r$.  Well-posedness of this problem requires $p(x) \geq p_{min} > 0 \textrm{ and } q(x) \geq q_{min} \geq 0$ 
Furthermore, for convenience, it is assumed that $p,q \in C([x_l,x_r])$ and $f \in L^2([x_l,x_r])$. These conditions are sufficient to show the existence of a unique solution to the weak form (\cite{levitan1975introduction}). 

\subsection{Functional minimization}

Motivated by the intractability of direct integration of the differential equation (\ref{eq_strongform}), it is often convenient to employ functional minimization techniques. Calculus of variations can be used to observe that the minimization of the functional  \eqref{eq_energyform} leads to the strong form described in Eq \eqref{eq_strongform}.
\begin{equation}\label{eq_energyform}
    \Pi\left[u\right] = \int_{x_l}^{x_r} \left( \frac{1}{2}pu'^2 + \frac{1}{2}qu^2 -fu \right)dx
\end{equation}
Square integrability of $u$  and its first derivative are required in this definition of $\Pi\left[u\right]$. The implication is that the minimizing solution, $u$ lies in the Sobolov space $H^1(\left[x_l, x_r\right])$. 
A discrete problem is obtained by using a finite basis for the solution defined in Eq \eqref{eq_basis} which satisfies the Dirichlet boundary conditions. The admissible choices of $\mathbf{a}=(a_0, a_1, ..., a_N)$ satisfy $u_N(x_d)=u_d$ where $x_d$ is a Dirichlet boundary and $u_d$ is the prescribed value at that point. This approximation reduces the infinite dimensional functional minimization problem to finite dimensions. The approximated functional $\Pi_N$ is entirely determined by the representation of $u$ in the finite basis as shown in Eq (\ref{eq_discretenergy}). 
It is worth observing that the choice of $\phi_i(x)$ is such that $\phi_i \in H^1(\left[ x_l, x_r  \right])$ i.e. for any  $u_N \in V_N = span\lbrace\phi_1, \phi_2, ...,\phi_N \rbrace \subseteq H^1(\left[ x_l, x_r  \right])$. Additionally the proper inclusion, $V_i \subseteq V_{i+1}$, guarantees convergence of the solution with increasing $N$.
\begin{equation}\label{eq_basis}
    u_N(x) = \sum_{i=0}^N a_i \phi_i(x)
\end{equation}

\begin{gather}
    \Pi_{N}\left[a_0, ...,a_r,.. ,a_N \right] = \int_{x_l}^{x_r}  \frac{p}{2}\left(\sum_{i=0}^{N} a_i \phi'_i\right)^2 \nonumber\\+ \frac{q}{2}\left(\sum_{i=0}^{N} a_i \phi_i\right)^2 -f\left(\sum_{i=0}^{N} a_i \phi_i\right) dx \label{eq_discretenergy}
\end{gather}

As the solution is completely determined by the variable $\mathbf{a}$, the functional minimization of Eq \eqref{eq_discretenergy} is reformalized as Eq \eqref{eq_basol} where $\mathbf{a}^{b.a.}$ refers to the coefficients of best approximation of solution, $u_N$, in the subspace $V_N$

\begin{equation}\label{eq_basol}
    \mathbf{a}^{b.a.} = \argmin_{\mathbf{a}} \Pi_{N}(\mathbf{a})
\end{equation}

\subsection{Finite Element approximation}
Finite element basis provides a popular choice of compactly supported shape functions. For the purpose of simplicity, `tent/hat functions' (defined in Eq \eqref{eq_hatfunction}) are used in this work. The domain is split into $N$ elements with $N+1$ nodes. The generalization to higher order families of piecewise-continuous basis functions is immediate but is omitted for brevity. 

\begin{equation}\label{eq_hatfunction}
     {\phi}_i(x) = \left\lbrace
     \begin{array}{ll}
     0, & x < x_{i-1},\\
     (x - x_{i-1})/(x_{i} - x_{i-1}),
     & x_{i-1} \leq x < x_{i},\\
     1 -
     (x - x_{i})/(x_{i+1} - x_{i}),
     & x_{i} \leq x < x_{i+1},\\
     0, & x\geq x_{i+1}{\thinspace .} 
     \end{array}
     \right.
\end{equation}

The usage of compact basis further reduces the complexity by reducing the integration over the whole domain to a summation of integration over smaller elements. It is shown in section \ref{sec_Graph_model} that this choice of shape functions lead to a relatively sparse graph. It simplifies the computation by reducing the size of the graph optimization problem. The simplified form of $\Pi$ specialized for the hat-functions is presented in Eq \eqref{eq_functional_elementsum}.

\begin{gather}
    \Pi_N(\mathbf{a})  
    =\sum_{i=1}^{N} a_{i-1}^2 \left( \int_{x_{i-1}}^{x_{i}} \frac{p}{2}  \phi'^2_{i-1} + \frac{q}{2}  \phi^2_{i-1} dx \right) \nonumber\\
    + a_{i}^2 \left( \int_{x_{i-1}}^{x_{i}} \frac{p}{2}  \phi'^2_{i} + \frac{q}{2}  \phi^2_{i} dx \right)\nonumber\\
    + a_{i-1}a_{i} \left(\int_{x_{i-1}}^{x_{i}} p\phi'_{i-1} \phi'_{i} + q\phi_{i-1} \phi_{i} dx\right)\nonumber \\\label{eq_functional_elementsum}
    - a_{i-1}\left(\int_{x_{i-1}}^{x_{i}} f\phi_{i-1}dx\right) -a_{i}\left(\int_{x_{i-1}}^{x_{i}} f\phi_{i}dx\right)
\end{gather}

This form of $\Pi$ promotes modularity in computation
and allows expressing the functional as 
\begin{equation}\label{aisi}
   \Pi_N = \sum_{i=1}^{N} \mathbf{A_i}.\mathbf{S_i}
\end{equation}
where vectors $\mathbf{A_i}\equiv \mathbf{A_i}(a_{i-1},a_i)$ and $\mathbf{S_i}\equiv \mathbf{S_i}(p,q,f)$ are defined for each element in \eqref{eq_functionalIP}. The vector $\mathbf{S_i}$ is independent of state $\mathbf{a}$ and is therefore only computed once in the whole procedure. 
\begin{gather}
\mathbf{A_i} = \Big[a^2_{i-1} \quad,\quad a^2_{i} \quad,\quad a_{i-1}a_{i} \quad,\quad a_{i-1}\quad,\quad a_{i} \Big]^T \nonumber\\
\mathbf{S_i} = \Bigg[  \int_{x_{i-1}}^{x_{i}} \frac{p}{2}  \phi'^2_{i-1} + \frac{q}{2}  \phi^2_{i-1} dx  \quad,\quad   \int_{x_{i-1}}^{x_{i}} \frac{p}{2}  \phi'^2_{i} + \frac{q}{2}  \phi^2_{i} dx  \quad, \nonumber \\
\int_{x_{i-1}}^{x_{i}} p\phi'_{i-1} \phi'_{i} + q\phi_{i-1} \phi_{i} dx \quad,\quad -\int_{x_{i-1}}^{x_{i}} f\phi_{i-1}dx \quad,\nonumber \\
\quad -\int_{x_{i-1}}^{x_{i}} f\phi_{i}dx \Bigg]^T\label{eq_functionalIP}
\end{gather}

\vspace{5pt}
\hrule
\vspace{5pt}

\begin{center}
    Example
\end{center}
Consider the differential equation  with boundary conditions $u(0)=0$ and $u(1) = 1$. 
\begin{gather*}
    \frac{d^2 u }{dx^2} = 0 \qquad 0<x<1
\end{gather*}
Functional: 
\begin{eqnarray*}
\Pi[u] = \frac{1}{2}\int_0^1 u'^2 dx
\end{eqnarray*}
For simplicity, consider a grid with a uniform mesh of 2 elements and 3 nodes: 
\begin{figure}[H]
\centering
\includegraphics[width=0.4\textwidth]{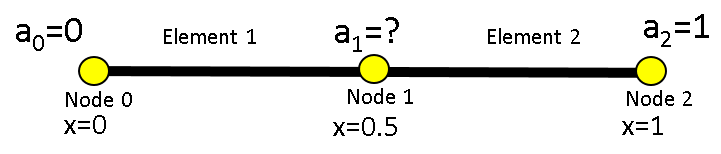}
\end{figure}
Using linear interpolants for the elements,
\begin{eqnarray*}
u(x) = \left\{\begin{matrix}
a_0 (1 - 2x)+ a_1 (2x) & 0<x\leq 0.5\\\\ 
a_1 (2 - 2x)+ a_2 (2x-1) & 0.5<x\leq 1\\
\end{matrix}\right.
\end{eqnarray*}
The functional with the FE discretization: 
\begin{eqnarray*}
\Pi_N(\textbf{a}) = (a_0-a_1)^2 + (a_1-a_2)^2
\end{eqnarray*}
Modular representation of functional ($\Pi_N = \mathbf{A_1}.\mathbf{S_1} + \mathbf{A_2}.\mathbf{S_2}$): 
\begin{gather*}
   \mathbf{A_1} = \Big[a^2_{0} \quad,\quad a^2_{1} \quad,\quad a_{0}a_{1} \quad,\quad a_{0}\quad,\quad a_{1} \Big]^T \\
   \mathbf{A_2} = \Big[a^2_{1} \quad,\quad a^2_{2} \quad,\quad a_{1}a_{2} \quad,\quad a_{1}\quad,\quad a_{2} \Big]^T \\ 
   \mathbf{S_1} = \mathbf{S_2}  =  \Big[1 \quad,\quad 1 \quad,\quad -2 \quad,\quad 0 \quad,\quad 0 \Big]^T 
\end{gather*}
\vspace{5pt}
\hrule
\vspace{5pt}

\section{Graph Coloring Problem}\label{sec_Graph_model}

Quantum annealing methods are tailored to find the lowest energy states in an Ising system defined in Eq \eqref{eq_Ising}. The Ising hamiltonian defines a binary graph coloring problem with each vertex of graph or qubit labeled as $+1$ or $-1$. The value of the qubits determine the free variable, in this case, $\mathbf{a}$. The parameters $H_i$ and $J_{ij}$ are defined such that the Ising hamiltonian, for a labeling representing the state, $\mathbf{a}$, corresponds to the functional $\Pi_N(\mathbf{a})$. These problems, namely, the representation of state and estimation of parameters are addressed in this section.  

\subsection{Representation of State}
Representation of a functional in terms of continuous variables is not feasible on quantum architectures. Due to this limitation, the values of each $a_i$ ($i^{th}$ component of $\mathbf{a}$) are chosen from a finite set of values based on the labeling of qubits. The representation presented here permits 3 possible values of $a_i$ at each node. In particular, for each node (indexed `$i$'), the state ($a_i$) is exactly determined by the labeling of qubits $q^i_1$, $q^i_2$ and $q^i_3$ with the $i^{th}$ node taking values in the set $\{ v_{i_1}, v_{i_2}, v_{i_3} \}$. Eq \eqref{eq_evaltestfunc} defines a mapping $(q^i_1,q^i_2,q^i_3)\rightarrow a_i$ as tabulated in Table \ref{table_qub2statemap}. It is observed that the mapping results in $a_i \in \{v_{i_1}, v_{i_2}, v_{i_3} \}$ when two qubits are labeled $-1$ and one qubit is labeled $+1$. Next it shown that the Ising parameters can be manipulated to make these labelings energetically favourable, thereby eliminating the occurrence of undesirable labels. 

\begin{equation}\label{eq_evaltestfunc}
    a_i = \sum_{j=1}^3 v_{i_j} \frac{q^i_j +1}{2}
\end{equation}

\begin{table}[] 
\begin{tabular}{|c|c|}
$(q^i_1,q^i_2,q^i_3)$   & $a_i$ \\\hline
$(1  , 1  , 1  )$& $v_{i_1}+v_{i_2}+v_{i_3}$ \\
$(1  , 1  , -1 )$& $v_{i_1}+v_{i_2}$ \\
$(1  , -1 , 1  )$& $v_{i_1}+v_{i_3}$ \\
$(1  , -1 , -1 )$& $v_{i_1}$ \\
$(-1 , 1  , 1  )$& $v_{i_2}+v_{i_3}$ \\
$(-1 , 1  , -1 )$& $v_{i_2}$ \\
$(-1 , -1 , 1  )$& $v_{i_3}$ \\
$(-1 , -1 , -1 )$& $0$
\end{tabular}
\caption{The mapping from qubits to state $a_i$ at node}
\label{table_qub2statemap}
\end{table}

\vspace{5pt}
\hrule
\vspace{5pt}

\begin{center}
    Example (Continued)
\end{center}
In general, the set $\{v_{i_1}, v_{i_2}, v_{i_3}\}$ is different for each node. However, for simplicity, consider the same set of admissible states for all three nodes given by $\{v_{i_1}, v_{i_2}, v_{i_3}\} \equiv \{0,0.5,1\}$. Each node is defined by three qubits as follows:

\begin{figure}[H]
\centering
\includegraphics[width=0.4\textwidth]{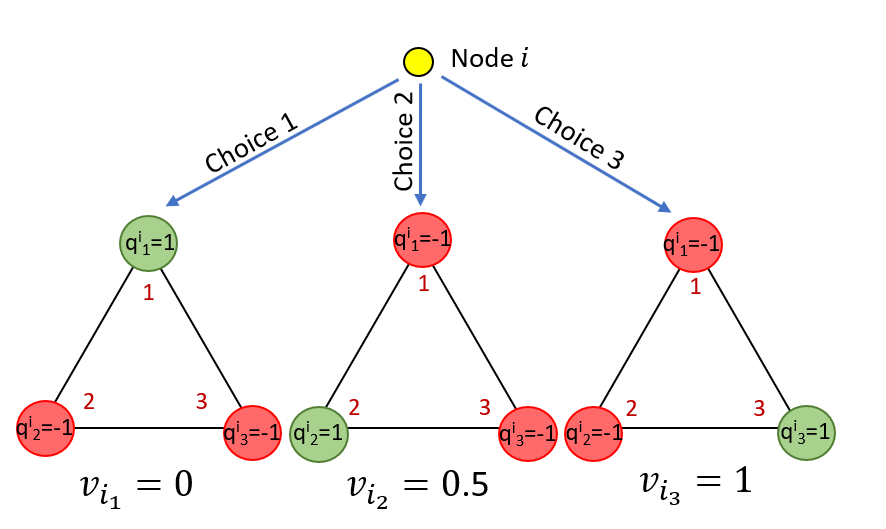}
\end{figure}

We know that the three qubits each defining the solution at the first and last nodes should take up choices 1 and 3, respectively, due to boundary conditions. The choice for the second node is to be solved.

\vspace{5pt}
\hrule
\vspace{5pt}
\subsection{Parameter Estimation}
To promote modularity, the graph representation is decomposed into two component subgraphs, namely, nodal graph and element graph. Each node and element of the FE discretization is endowed with a node graph and element graph, respectively. This allows to refine the mesh by extending the graph. 

\begin{figure}[hbt!]
\centering
\includegraphics[width=0.4\textwidth]{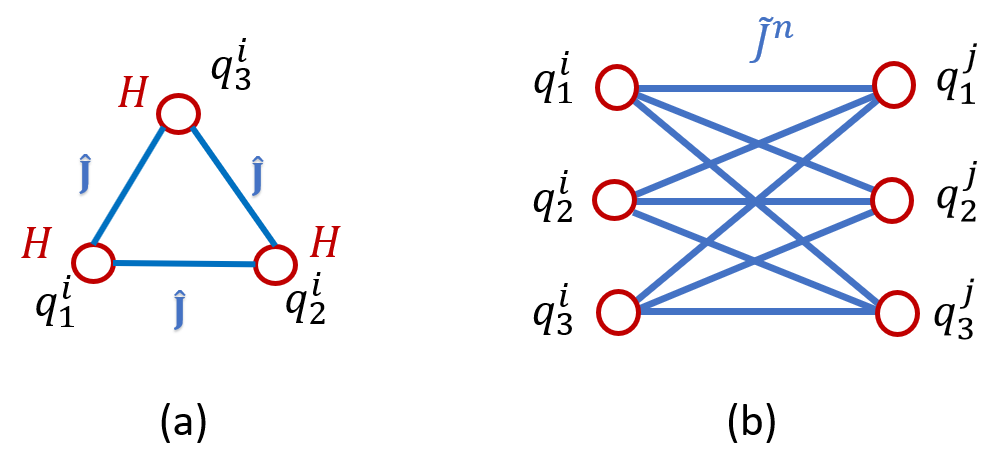}
\caption{Connectivity of (a) nodal graph (b) element graph.}
\label{fig_elementsubgraph}
\end{figure}

\subsubsection{Nodal Graph}
\begin{figure*}[hbt!]
\centering
\includegraphics[width=0.8\textwidth]{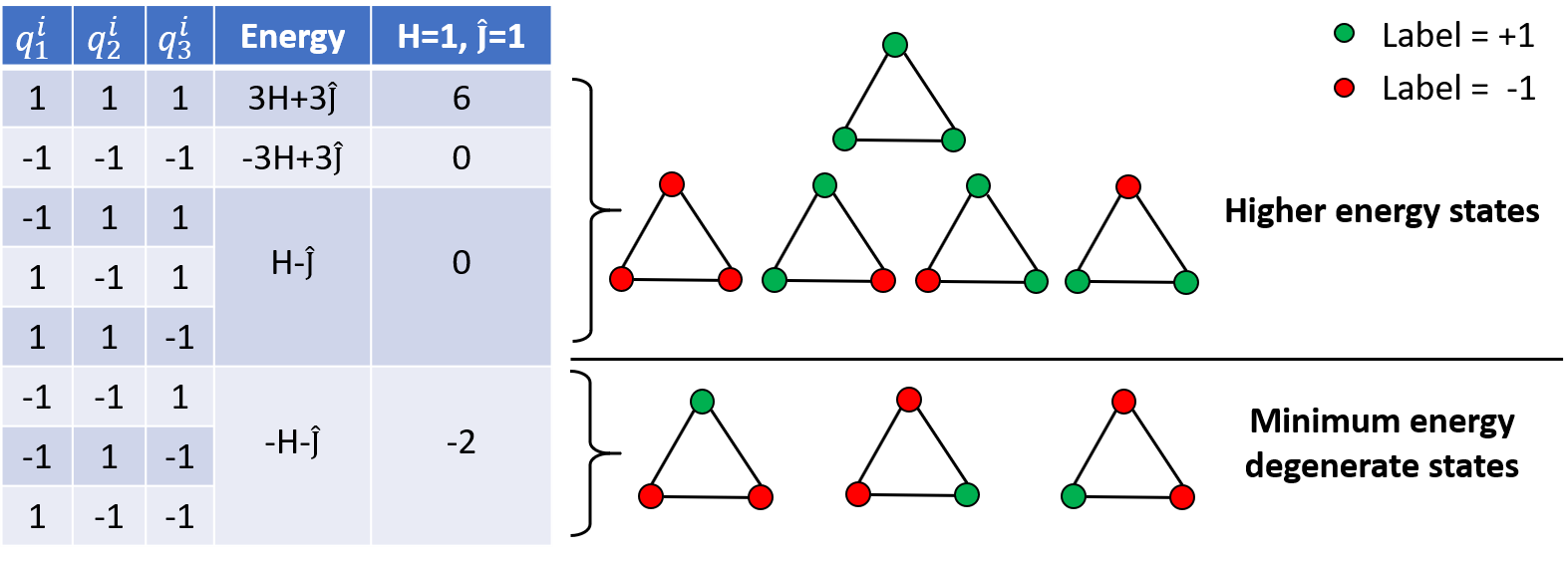}
\caption{Self interaction and site-site interaction parameters for nodal subgraph.}
\label{fig_nodalgraph}
\end{figure*}

The nodal graph is given by a fully connected graph with three vertices representing the three qubits of the FE node. The nodal graph ensures that the energy minimizing states of the Ising hamiltonian corresponds to state $\mathbf{a}$ with favorable choice of $a_i \in \{ v_{i_1}, v_{i_2}, v_{i_3} \}$ with equal probability. As mentioned earlier, the set of favorable labeling of qubits at a node is given by $\{Q_1, Q_2, Q_3\} \equiv \{ (1,-1,-1),(-1,1,-1),(-1,-1,1) \}$. Since each of the three labelling is equally likely in the absence of any functional minimization, it is expected that the same value of the coupling strength ($\hat{J}$) for each connection and the field strength ($H$) for each node is used. A choice of $\hat{J}$ and $H$ that fulfill these conditions is presented in Fig \ref{fig_nodalgraph}. Here, all the field and interaction terms for the nodal graph are given a value of one. The application of the Dirichlet boundary condition is also done by augmenting the field strength of the nodal graph. For example, by switching the field term $H$ corresponding to the second qubit $q_2$ of a boundary node `b' to -1 forces a lower value of the functional for the boundary node state of $( -1, +1, -1)$, which corresponds to the solution $v_{b_2}$. This allows us to encode the value at the boundary to be $v_{b_2}$.

\subsubsection{Element Graph}

The element graph is used to make the energy of minimizing states of graph correspond to the value of functional $\Pi_N$ of the continuous problem. Each element graph encodes the contribution of the respective element to the functional. Since the contribution of each element is dependent on the values at the nodes of the element, the element graph is constructed by connecting the vertices of neighbouring nodes. In particular, the site-site interaction in the $n^{th}$ element graph can be estimated as a matrix, $\widetilde{J}^n$, where $(\widetilde{J}^n)_{kl}$ represents the coupling energy between qubits $q^i_k$ ($k^{th}$ qubit of $i^{th}$ node) and $q^j_l$ ($l^{th}$ qubit of $j^{th}$ node) with $i,j$ being the nodes of $n^{th}$ element. As shown in the previous section, the contribution of the $n^{th}$ element towards the functional, based on the choice of a compact basis function, is evaluated as $\mathbf{A_n}.\mathbf{S_n}$. The elements of the vector, $\mathbf{A_n}\equiv \mathbf{A_n}(a_{i},a_{j})$ can therefore, take nine ($3\times 3$) possible values based on the values of $(a_{i},a_{j})$. For a particular choice of labeling of qubit the Ising energy of element graph is estimated as $E = \sum_{k=1}^3 \sum_{l=1}^3 (\widetilde{J}^n)_{kl} q^i_k q^j_l$. When the labeling is chosen appropriately (each node has two `-1' and one `+1' label), this energy equals to the value of functional for corresponding state, $\mathbf{a}$, as shown in Eq \eqref{eq_estimatesitesite}. This relation can be used to estimate $\widetilde{J}^n$ by solving a set of nine independent linear equations presented. It is important observation that the independence of these set of equations relies on the fact that for any node, $v_{i_k}\neq v_{i_l}$ for $k\neq l$. Additionally, the energy of the element graph breaks the symmetry between the states that minimize the energy of nodal graph, however, the values of $\widetilde{J}^n$ should be judiciously scaled (uniformly along all elements) such that energy of unfavourable states remain high.  

\begin{equation}\label{eq_estimatesitesite}
    \sum_{k=1}^3 \sum_{l=1}^3 (\widetilde{J}^n)_{kl} q^i_k q^j_l = \mathbf{A_n}(a_{i},a_{j}).\mathbf{S_n}
\end{equation}

\vspace{5pt}
\hrule
\vspace{5pt}

\begin{center}
    Example (Continued)
\end{center}
A single element with two nodes admits to the following connectivity: 
\begin{figure}[H]
\begin{center}
\includegraphics[width=0.5\textwidth]{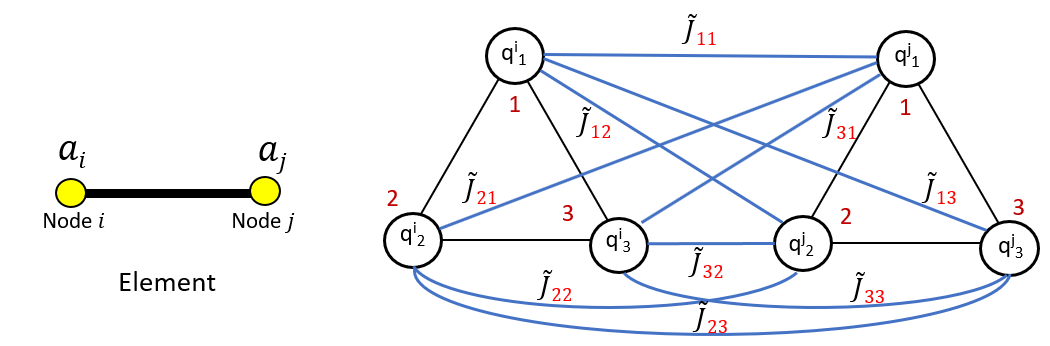}
\end{center}
\end{figure}
The estimated parameters reflect the contribution of element to the functional for a given choice of labeling: 

Sample 1: \\
\begin{figure}[H]
\begin{center}
    \includegraphics[width=0.5\textwidth]{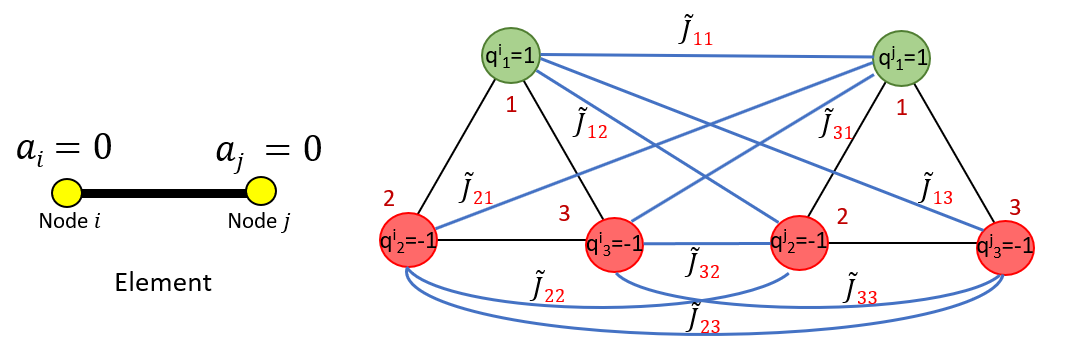}
\end{center}
\end{figure}
In the above Figure, both nodes take up choice 1 ($a_i=a_j=0$). The interaction energy for qubits: $E = \widetilde{J}_{11} - \widetilde{J}_{12} - \widetilde{J}_{13} - \widetilde{J}_{21}  +\widetilde{J}_{22} + \widetilde{J}_{23} - \widetilde{J}_{31} + \widetilde{J}_{32} + \widetilde{J}_{33} = (a_i-a_j)^2 = 0 $\\
Sample 2: \\
\begin{figure}[H]
\begin{center}
\includegraphics[width=0.5\textwidth]{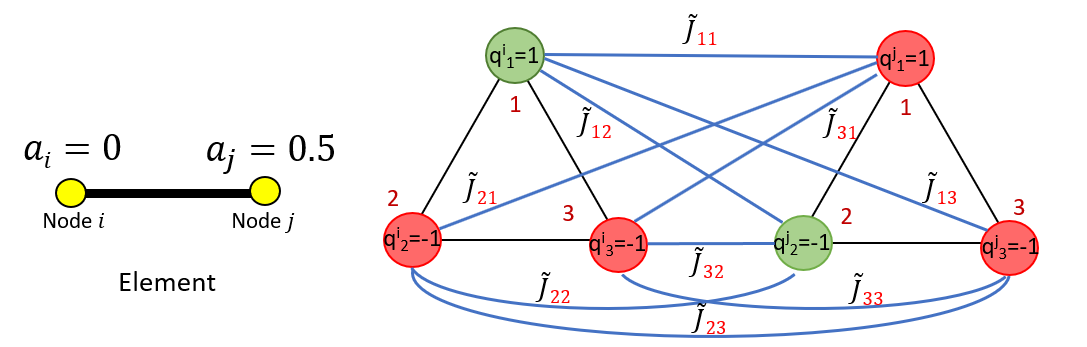}
\end{center}
\end{figure}
In the above Figure, node i takes up choice 1 ($a_i=0$), while node j takes up choice 2 ($a_j=0.5$). The interaction energy for qubits: $E = -\widetilde{J}_{11} + \widetilde{J}_{12} + \widetilde{J}_{13} + \widetilde{J}_{21}  -\widetilde{J}_{22} - \widetilde{J}_{23} - \widetilde{J}_{31} + \widetilde{J}_{32} + \widetilde{J}_{33} = (a_i-a_j)^2 = 0.25$ \\

Collectively solving the equation for all 9 such possibilities (as shown in Eq \eqref{eq_estimatesitesite}).
\begin{gather}
    \begin{bmatrix}
 +1& -1&  -1& -1& +1& +1& -1& +1& +1\\
 -1& +1&  +1& +1& -1& -1& -1& +1& +1\\
 -1& +1&  +1& -1& +1& +1& +1& -1& -1\\
 -1& +1&  -1& +1& -1& +1& +1& -1& +1\\
 +1& -1&  +1& -1& +1& -1& +1& -1& +1\\
 +1& -1&  +1& +1& -1& +1& -1& +1& -1\\
 -1& -1&  +1& +1& +1& -1& +1& +1& -1\\
 +1& +1&  -1& -1& -1& +1& +1& +1& -1\\
 +1& +1&  -1& +1& +1& -1& -1& -1& +1
\end{bmatrix}
\begin{bmatrix}
\widetilde{J}^n_{11}\\ 
\widetilde{J}^n_{12}\\ 
\widetilde{J}^n_{13}\\ 
\widetilde{J}^n_{21}\\  
\widetilde{J}^n_{22}\\ 
\widetilde{J}^n_{23}\\ 
\widetilde{J}^n_{31}\\ 
\widetilde{J}^n_{32}\\
\widetilde{J}^n_{33}
\end{bmatrix} \nonumber\\
= \begin{bmatrix}
(v_{i_{1}}-v_{j_{1}})^2\\
(v_{i_{2}}-v_{j_{1}})^2\\
(v_{i_{3}}-v_{j_{1}})^2\\
(v_{i_{1}}-v_{j_{2}})^2\\
(v_{i_{2}}-v_{j_{2}})^2\\
(v_{i_{3}}-v_{j_{2}})^2\\
(v_{i_{1}}-v_{j_{3}})^2\\
(v_{i_{2}}-v_{j_{3}})^2\\
(v_{i_{3}}-v_{j_{3}})^2 \nonumber
\end{bmatrix}\label{eq_estimatesitesite2}
\end{gather}

\begin{equation*}
\widetilde{J}^1 =  \widetilde{J}^2 =   \begin{bmatrix}
    0.1250   &  0.3750  &    0.3750\\
    0.3750   &  0.5000  &   0.3750\\
    0.3750   & 0.3750  &  0.1250
\end{bmatrix}
\end{equation*}

The above parameters will exactly reproduce the functional in the interaction term. The boundary conditions are enforced, by setting self interaction term for qubits $q^0_1$, $q^2_3$ to $H=-1$. This locks the state at the $1^{st}$ boundary node as $a_0 = v_{0_1} = 0$ and at the $2^{nd}$ boundary node as $a_2 = v_{2_3} = 1$. Energy minimization of the resulting Ising hamiltonian gives $a_1 = v_{1_2}= 0.5$, which is the exact solution for the discretized problem. 
\vspace{5pt}
\hrule
\vspace{5pt}

\begin{figure*}[hbt!]
\centering
\includegraphics[width=0.7\textwidth]{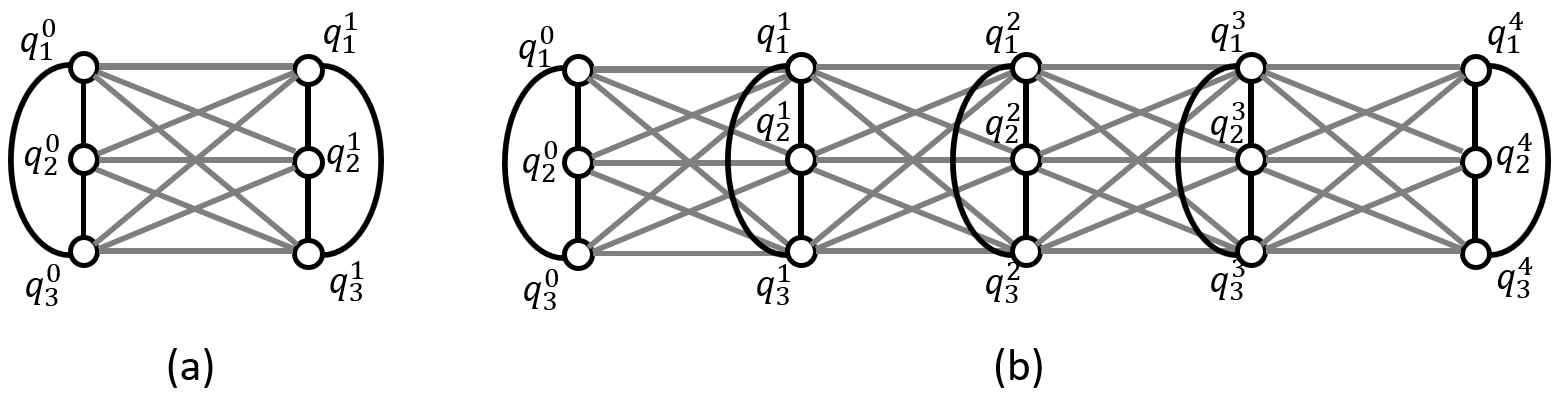}
\caption{Assembled graph for a domain discretized with (a) 1 element (b) 4 elements.}
\label{fig_assembly}
\end{figure*}

The process of the graphical representation of the discretized functional using the nodal and element graphs is referred to as ``Assembly". Each node and element is endowed with a nodal and element graph, respectively. The effective site-site interaction energy is estimated by summing the nodal coupling strength, $\hat{J}$, over all nodes, and element coupling strength, $\widetilde{J}$, over all elements. Due to the nature of discretization, $N$ element graphs and $N+1$ nodal graphs are required for representing an $N$-element discretization of the domain. The assembled graph, from here on, is referred to as the logical graph. Connectivity of logical graphs for one-element and four-element discretization is presented in Fig \ref{fig_assembly}.

Two fundamental issues in this approach are addressed next using the box algorithm. Firstly, the choices at a node $\{ v_{i_1}, v_{i_2}, v_{i_3} \}$ were set in stone during initialization. The box algorithm makes this choice flexible. Secondly, as the number of nodes increase, three choices are insufficient. The number of qubits needed at a node must increase to make more choices available. Box algorithm, however, only requires three qubits per node for any level of discretization.

\section{Box Algorithm}\label{sec_Box}

In this section, an iterative procedure is developed to minimize the functional, $\Pi_N$, using the graph coloring representation discussed in the previous section. For a particular choice of $\{v_{i_1},v_{i_2},v_{i_3}\}$, defined as Eq \eqref{eq_centerbound}, the possible values of the state $a_i$ at the $i^{th}$ node are specialized to the set $\lbrace u_i^c - r , u_i^c , u_i^c + r \rbrace$, i.e., 

\begin{equation}\label{eq_centerbound}
    v_{i_j} = u_i^c + r(j-2)
\end{equation}

\begin{figure*}[hbt!]
\centering
\includegraphics[width=0.9\textwidth]{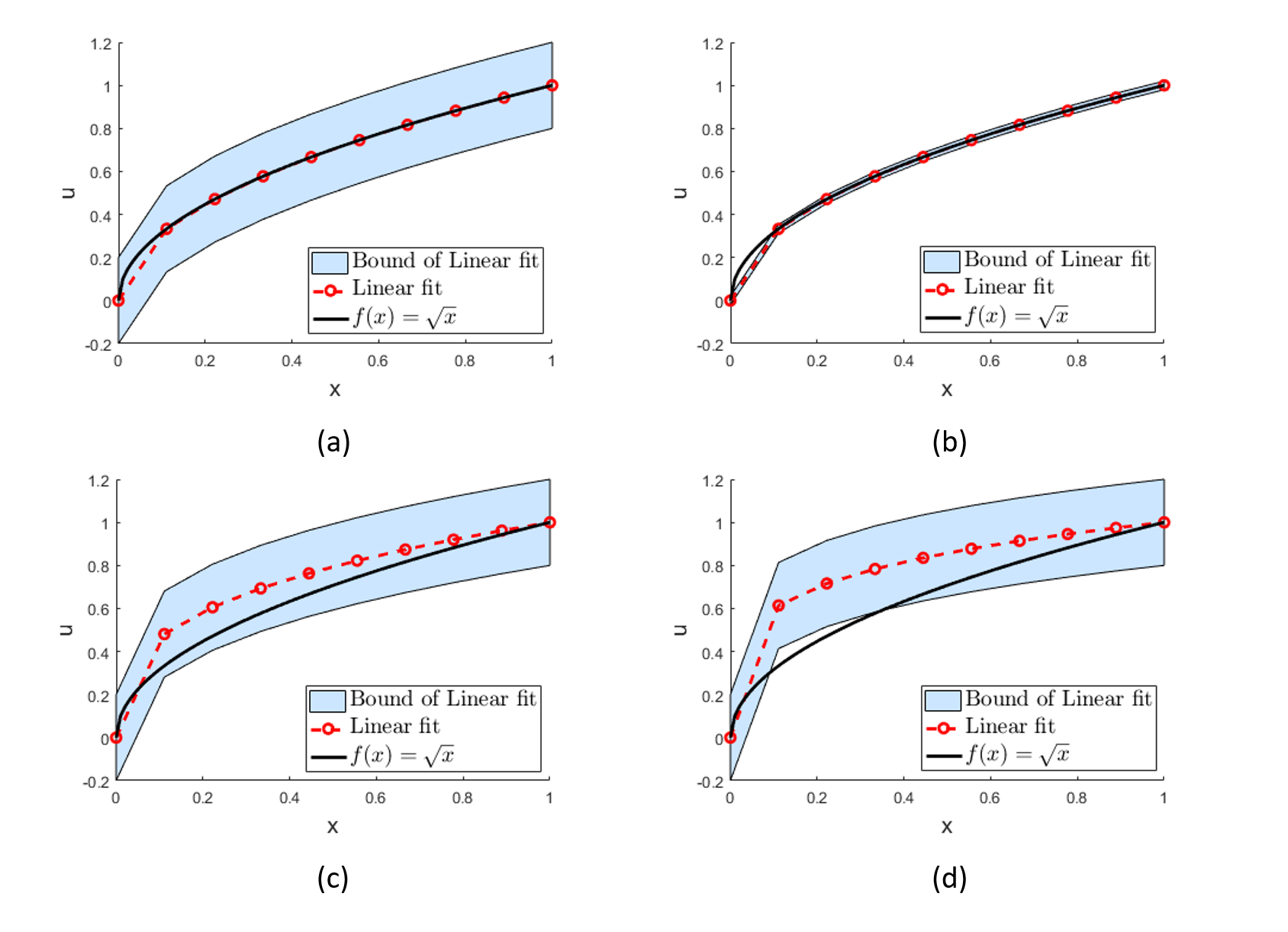}
\caption{Approximation of $\sqrt{x}$ function using boxed domain: (a) Exact fit with a slack of $0.2$ (Loose fit) (b) Exact fit with a slack of $0.02$ (tight fit) (c) inexact fit but bounded in a box size of $0.2$ (d) inexact fit and unbounded by a slack of $0.2$.}
\label{fig_box}
\end{figure*}

\begin{figure}[hbt!]
\centering
\includegraphics[width=0.35\textwidth]{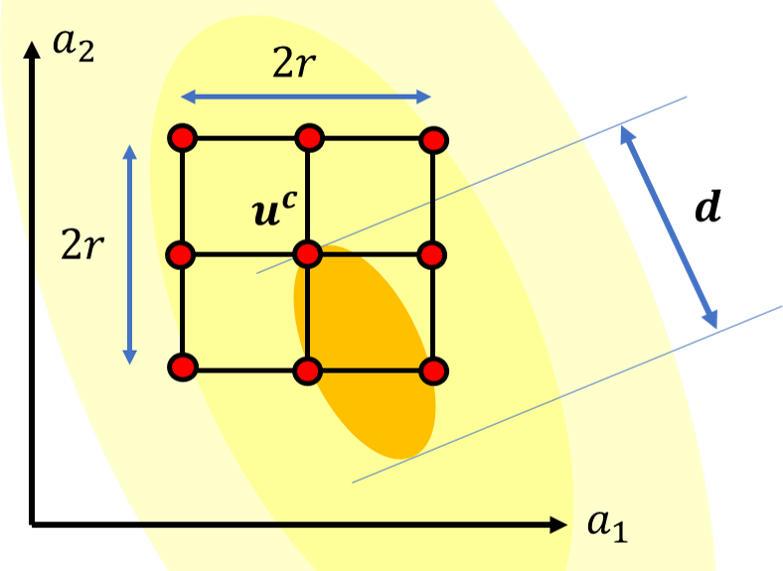}
\caption{ An illustration of a two-node approximation in $a_1-a_2$ vector space with contours plot of the functional, $\Pi_2( a_1,a_2)$ and a representative box with center at $\mathbf{u^c}$ and a box size of $r$.}
\label{fig_box2}
\end{figure}

The quantities, $\mathbf{u^c}= ( u_0^c, u_2^c,..., u_{N}^c )$ and $r$  are hereafter referred to as box center and the slack variable, respectively. The intention is to approximate functions using the box center while the slack variable provides a bound on this approximation. The precise meaning of this bound is presented later in this section. A linear approximation of $f(x) = \sqrt{x}$ using ten nodes is presented in Fig \ref{fig_box} for different box centers and the slack variable (which can be interpreted as the box size). The function, $f(x)$, is approximated as $\mathbf{u^c}$ at the nodes with linear interpolation in between the nodes. The blue region describes the possible value of the interpolation if the value at any node is perturbed within the range of $\pm r$. In Fig \ref{fig_box}(a), an exact approximation of the function at the nodes is presented with a slack variable of $0.2$. In (b) the same approximation with a slack variable of $0.02$ is presented. The same approximation is given in the two cases, but the bound on nodal values of (b) is tighter than (a). In part (c), the approximation is not exact, however, it lies within the bounds. In part (d), the approximation is neither exact nor within the bound. In the context of the vectorial representation of the coefficients, $\mathbf{a}$, these bounds are represented as $3^N - 1$ points on the surface of a box, defined as, $||\mathbf{a}- \mathbf{u^c}||_\infty = r$. An illustration for the vectorial representation of a two nodes element is presented in Fig \ref{fig_box2}.  The solution is sampled from a $3\times3$ grid in the $a_1-a_2$ vector space.\\

In the discrete setting, the solution to the differential equation can be equivalently reduced to minimization of a function of the form:  $\mathbf{a^T M a}$ where $\mathbf{M}$ is some positive definite matrix. The vector $\mathbf{a}$ takes value in one of the $3^N$ possibilities. The minimizer (not necessarily unique) is given by Eq \eqref{eq_isingmin}. The solution, $\mathbf{a}^{min}$, need not coincide with the best approximation solution, $\mathbf{a}^{b.a.}$ of the continuous problem. In the illustration presented in Fig \ref{fig_box2}, the center is depicted as the solution ($\mathbf{a}^{min} = \mathbf{u}^c$), the minimum is then contained within the elliptic region of the contour with $\mathbf{a}^{min}$ on the edge. Geometrically, this gives $||\mathbf{a}^{min}- \mathbf{a}^{b.a}|| \leq d \leq \sqrt{2}r(1+\lambda_{max}/\lambda_{min})$ where $\lambda_{max}$ and $\lambda_{min}$ are the maximum and the minimum non--zero eigenvalues of $M$, respectively. This suggests that as the box size decreases, the corresponding $\mathbf{u}^c$ approaches to the best approximation solution of $u$. This argument is extended to $n$ dimensions with the bound given in Eq \eqref{eq_bound}. 

\begin{equation}\label{eq_isingmin}
     \mathbf{a}^{min} =   \argmin_{\substack{a_i\in \{u_i^c - r , u_i^c , u_i^c + r \}}} \mathbf{a^T M a}
\end{equation}

\begin{equation}\label{eq_bound}
    ||\mathbf{a}^{min}- \mathbf{a}^{b.a}|| \leq 2\left(1+(n-1)\frac{\lambda_{max}}{\lambda_{min}}\right)\frac{r}{\sqrt{n}}
\end{equation}

\subsection{Iterative Procedure}


In this section we present the details of the iterative procedure which locates the solution of the discretized problem, $\mathbf{a}^{min}$, and updates the box center and slack variable such that $\mathbf{u}^C$ approaches the solution of the continuous problem (in the sense of best approximation).  

The necessary information required for defining the functional is stored in the vector $\mathbf{S_i}$. It is computed once at the beginning of the procedure as the problem definition stage. The procedure is initiated with a guessed solution of the vector, $\mathbf{a}$, provided as a box centered at $\mathbf{u^c}$. The boundary nodes with Dirichlet boundary conditions are assigned the boundary value as the initial guess. The slack variable is initialized with an arbitrary scalar value. A better initial guess for $r$ is the one which bounds the solution in the box defined by $\mathbf{u^c}$. Such initial guesses require fewer iterations in comparison to arbitrary ones; however, starting with a good guess is not a necessary condition for convergence. The Ising parameters, $H$, $\hat{J}$ and $\widetilde{J}$ are estimated as discussed in section \ref{sec_Graph_model}.  \\

In this work, D-Wave's 2000Q processor is used. This processor has a Chimera-type structure with 2048 qubits and 6016 couplers \cite{dwavemanual}. A direct solution of the optimization problem by re-numeration of qubits is not possible as the assembled graph cannot be found in any subgraph of the physical graph, i.e., the processor. Therefore, it is required that the logical graph is mapped onto the physical graph via the process of embedding. This problem in itself is NP-hard and is not discussed here for brevity. The reader is referred to \cite{cai2014practical} for a discussion on this topic. The topology of the logical graph remains unchanged over the iterations. The search for embedding of a map is only conducted once, and in subsequent iterations, the self-interaction and the site-site interaction parameters are updated for the same embedding. 

The use of three qubits per node in this paper allows the D-wave system to search for a minimum over a space of $3^N$ solution vectors in a single run. In each iteration, the box center is translated to the energy minimizer, $\mathbf{a}^{min}$. This move is referred to as the translation step. In the case where the minimizing state is found at the center, the box size is reduced, and the search is continued with a smaller bound on error. This move is referred to as the contraction step. The complete procedure is presented in Algorithm \ref{algo}.

\begin{algorithm}[H]
\caption{Box Algorithm}\label{algo}
\begin{algorithmic}[1]
\State Problem definition: Calculate $\mathbf{S_i}$
\State Initialize $\lbrace u^c_i \rbrace$, $r$
\State Estimate $H$, $\hat{J}$ and $\widetilde{J}$
\State Find embedding: $\textrm{Logical graph} \xrightarrow{\text{embed}} \textrm{Physical graph}$
\While {$r>r_{\textrm{min}}$}
  \State Update $\widetilde{J}$ for current $(u_i^c,r)$
  \State Anneal for $\lbrace q^i_j \rbrace$
  \State Map to $\mathbf{a}^{min}$, $(\Pi_N)_{min}$
  \If{$(\Pi_N)_{min} < \Pi_N \left[\mathbf{u^c}\right]$}
   \State $u_i^c = \mathbf{a}^{min}$ (Translation step)
  \Else
   \State $r = \frac{r}{2}$ (Contraction step)
  \EndIf
\EndWhile
\State \textbf{end}
\end{algorithmic}
\end{algorithm}

\vspace{5pt}
\hrule
\vspace{5pt}

\begin{center}
    Example (Continued)
\end{center}
In the box algorithm, the set $\{v_{i_1},v_{i_2},v_{i_3}\}$ is constructed using the box center and the slack variable.
\begin{figure}[H]
\begin{center}
    \includegraphics[width=0.4\textwidth]{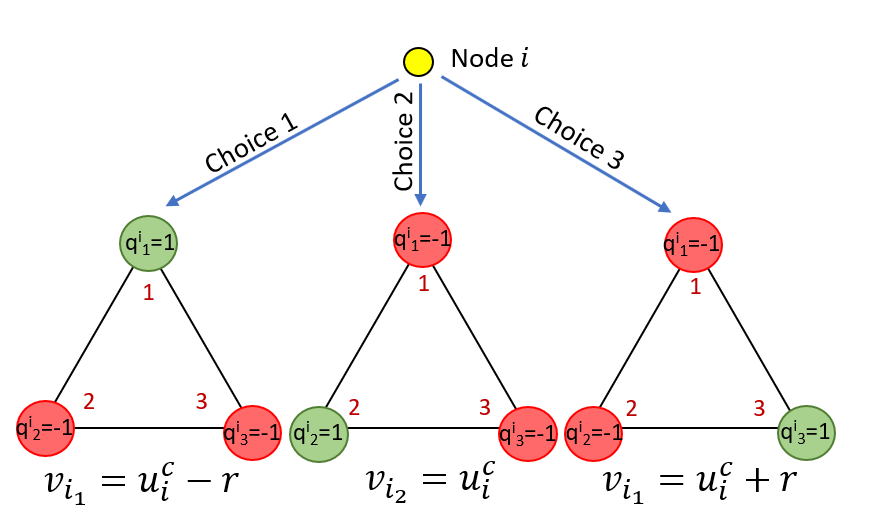}
\end{center}
\end{figure}
With the application of the boundary condition, the favourable labeling of qubits give following three choices in the solution ($\mathbf{I}$, $\mathbf{II}$, $\mathbf{III}$). 
\begin{figure}[H]
\begin{center}
    \includegraphics[width=0.4\textwidth]{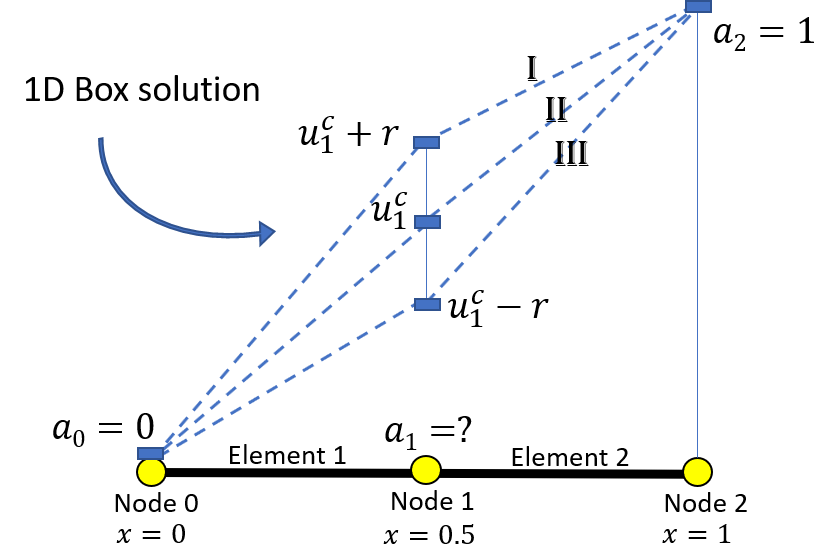}
\end{center}
\end{figure}
The values of  $u^c_1$ and $r$ are initialized arbitrarily. One of the solutions among $\mathbf{I}$, $\mathbf{II}$ or $\mathbf{III}$ is selected by the annealer. If the minimizer is found to be solution  $\mathbf{II}$ then the algorithm proceeds with the contraction step by halving the value of $r$. 
If solutions $\mathbf{I}$ or $\mathbf{III}$ are selected, then the algorithm proceeds with the translation step by setting the new box center to $u_1^c + r$ or $u_1^c - r$, respectively.
\vspace{5pt}
\hrule
\vspace{5pt}

\section{Results}\label{sec_results}
The deformation of a bar under axial loading is modelled using an equation of form \eqref{eq_strongform}. In particular, the deflection $(u)$ of a bar is related to the elastic stiffness, $(E)$, cross-sectional area $(A)$, and the applied body force, $(f)$ using Eq \eqref{eq_Elasticstrong}. The functional (Eq \ref{eq_energyform}) is referred to as the potential energy of the system. The corresponding discretized form of the potential energy for piece-wise linear $E$,$A$ and $f$ is calculated using Eq \eqref{eq_Elasticweak} where $E_i$,$A_i$ and $f_i$ represents the elastic stiffness, area and applied body force, respectively, at the center of the $i^{th}$ element.
\begin{eqnarray}\label{eq_Elasticstrong}
(EAu')' + f = 0
\end{eqnarray}
\begin{equation}\label{eq_Elasticweak}
    \Pi_N\left[u\right] = \sum_{i=1}^{N} \frac{N}{2} E_iA_i (a_{i}-a_{i-1})^2 - \frac{1}{2N}f_i (a_{i}+a_{i-1})
\end{equation}

\begin{figure*}
\centering
\includegraphics[width=1\textwidth]{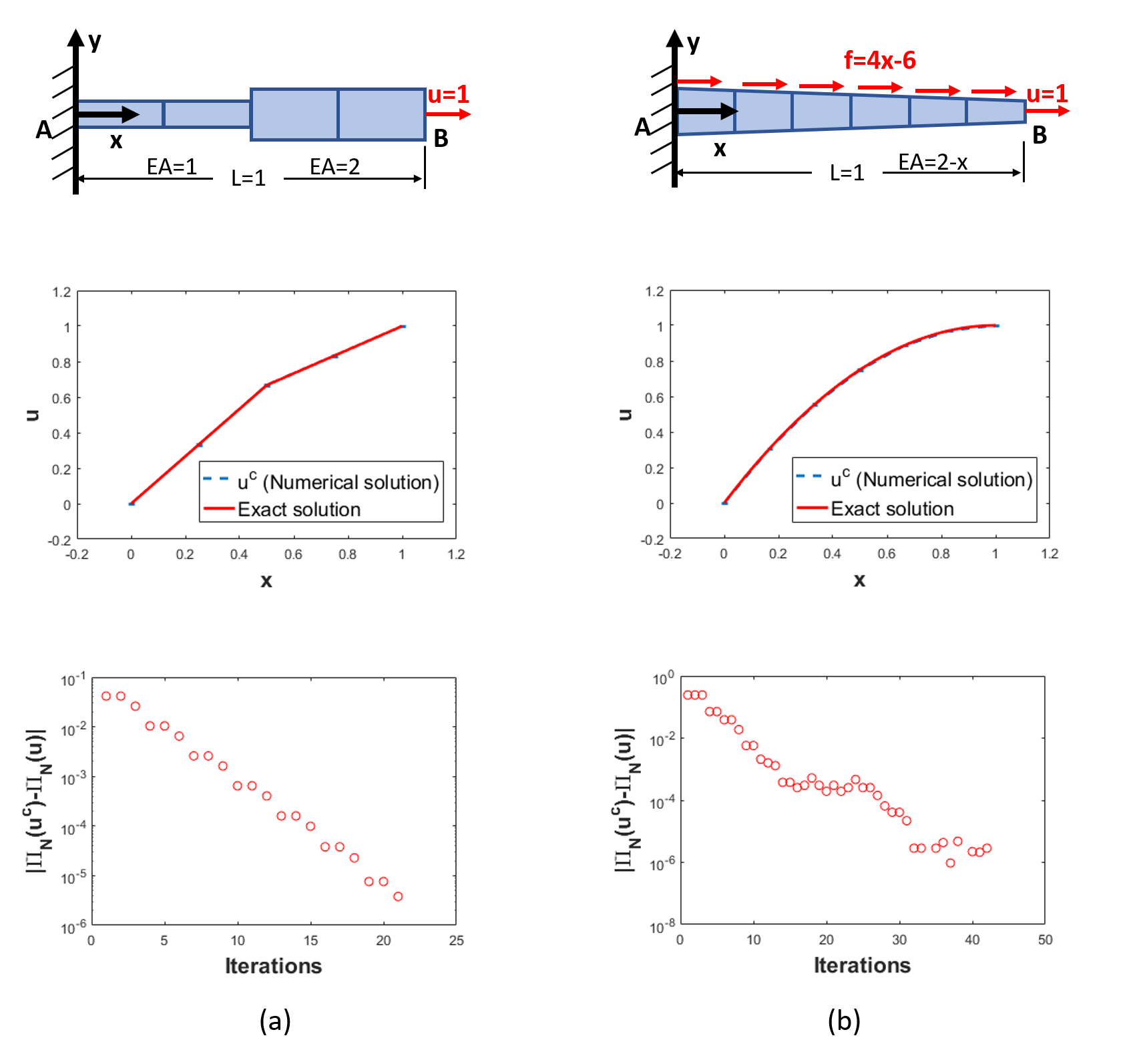}
\caption{Axial deformation of a bar with (a) a discontinuous cross-section with a tip displacement (b) a continuously varying cross section with a body force and a tip displacement.}
\label{fig_bars}
\end{figure*}
Two test cases are presented in Fig \ref{fig_bars}. In test case (a), a bar with a discontinuity in $EA$ is simulated. The body force is not applied in this case. A four-element discretization is used. The initial guesses are taken as $\mathbf{u^c} = \lbrace0,0.25, 0.5, 0.75, 1 \rbrace$ and $r=0.2$. The numerical solution is observed to approach the exact solution in this case. The convergence in the functional is also evident. 

In the test case (b), a bar with continuously varying $EA$ is simulated. A linearly varying body-force is supplied. A six-element discretization of the bar is used with $\mathbf{u^c} = \lbrace0,\frac{1}{6}, \frac{2}{6},\frac{3}{6},\frac{4}{6},\frac{5}{6},1 \rbrace$ and $r=0.2$. Based on the theory of finite element methods, the exact minimization of the energy in discretized space leads to a stiffer solution in comparison to the exact solution. It is observed that the numerical solution approaches the exact solution at nodes which is characteristic of finite element methods. Energy is also observed to be converging towards the finite element solution $u^{fem}$ in this case. The mismatch of $u$ within the element is expected to decrease with refinement in discretization.

Some implementation details on the D-wave architecture are relevant here. Although the mapping only requires three qubits per node, embedding of this graph into the Chimera graph produced an overhead of 9 qubits per node - constant over a range of discretizations. It is understandable since a complete graph of three qubits used to represent a node is not directly represented on the Chimera graph. In the future, the use of two qubits to represent a node can also be explored. While this ensures we still sample a large enough ($2^N$ vectors) solution space in a single run, we lose out information on the box center energy that is important for reducing the range of the slack variable. However, it is possible to compute the solution at the box center classically. The overhead of performing classical solutions can be offset by the fact that a two-qubit representation has smaller complete sub-graphs and is easier to embed in the physical graph. Another important task in quantum computing is error suppression. Quantum processors, unlike classical computers, do not have parity correction algorithms due to the no-cloning theorem. A compilation of popular methods for quantum error correction is presented in \cite{devitt2013quantum}. Energy re-scaling is one of the simpler approaches and is employed in this work. Here, in the estimation of $\widetilde{J}^n$, the energy was rescaled to ensure that the energy gap between feasible and unfeasible states is increased while maintaining a similar energy landscape. This step is a heuristic remedy for minimizing noise in quantum computation, and has no bearing in the theoretical convergence of the algorithm.

\section{Conclusions and Future Work}
Recent rapid developments in quantum annealers warrant further investigation into re-formulation of scientific computation problems as graph coloring problems. The use of quantum computing for solving differential equations has, to date, focused on the use of a gate--computing based QLSA algorithm. This algorithm attempts to sample from the solution space of the linear system of equations $Ax=b$. In the quantum annealer based algorithm described here, we do not solve the system of equations. We instead map the discretized version of the energy function of the differential equation to an Ising hamiltonian. The solution vector, $x$, is directly obtained as the ground state of the qubits. The algorithm has low memory requirements since the global matrix is not stored and the local matrices are encoded in the interaction weights of the Ising model. Further, the box algorithm allows mapping of up/down spin states of qubits in the ground state to rational numbers involved in the solution vector. 
Since we primarily solve the Ising model, the cost of computation is tied to the performance of the quantum annealer (\cite{mcgeoch2013experimental}). Within each iteration however, equation Eq \eqref{eq_estimatesitesite} is solved for each element, leading to at least $O(n)$ operations. 

We have shown that the box algorithm indeed guarantees convergence to the best approximation of the solution in the discretized space as the box size goes to zero. However, some improvements could be made to reduce the number of minimization runs. We could use the statistics of solutions that D-wave system returns from a single minimization run to drive the iteration process in an arbitrary direction. This data can also be used to heuristically develop `local' potential energy maps that can be used to identify larger step sizes for faster convergence. With future scaling up of quantum annealers up to millions of qubits, it will be possible to solve challenging engineering solid and fluid mechanics problems using quantum annealers.

\section{Acknowledgments}
The authors would like to acknowledge Universities Space Research Association (USRA) for providing access to the use of the D-Wave quantum computer in the USRA-NASA-Google quantum Artificial Intelligence Laboratory at NASA's Ames Research Center.

\bibliographystyle{apsrev}
\bibliography{references.bib}

\end{document}